\documentclass[conference]{IEEEtran}

\usepackage{graphicx}

\usepackage[T1]{fontenc}
\usepackage[utf8]{inputenc}

\usepackage{csquotes}
\usepackage{nicefrac}
\usepackage[textsize=scriptsize]{todonotes}
\usepackage{booktabs}
\usepackage{tikz}
\usetikzlibrary{shapes.geometric, positioning, quantikz}
\usepackage{hyperref}
\usepackage{amssymb}
\usepackage{amsthm}

\usepackage[font=footnotesize]{subcaption}

\usepackage{yquant}
\useyquantlanguage{groups}
\usepackage{pgfplots}

\usepackage[style=ieee, maxnames=6, minnames=1, date=year, doi=false,isbn=false,backend=biber]{biblatex}
\usepackage[binary-units=true, detect-all=true]{siunitx}
\usepackage{flushend}

\newtheorem{example}{Example}

\AtEveryBibitem{
  \clearname{editor}%
  \clearfield{series}%
  \clearfield{isbn}%
  \clearfield{issn}%
  \clearfield{volume}%
  \clearfield{number}%
  \clearfield{pages}%
  \clearfield{url}%
  \clearfield{doi}%
}

\DeclareFieldFormat
  [misc, book]
  {title}{\mkbibquote{#1}}

\begin{document}

\title{A Hybrid Classical Quantum Computing Approach to the Satellite Mission Planning Problem}

\author{
	\IEEEauthorblockN{Nils Quetschlich\IEEEauthorrefmark{1}\hspace*{1.5cm}Vincent Koch\IEEEauthorrefmark{2}\hspace*{1.5cm}Lukas Burgholzer\IEEEauthorrefmark{1}\hspace*{1.5cm}Robert Wille\IEEEauthorrefmark{1}\IEEEauthorrefmark{3}}
	\IEEEauthorblockA{\IEEEauthorrefmark{1}Chair for Design Automation, Technical University of Munich, Germany}
	\IEEEauthorblockA{\IEEEauthorrefmark{2}Technical University of Munich, Germany}
	\IEEEauthorblockA{\IEEEauthorrefmark{3}Software Competence Center Hagenberg GmbH (SCCH), Austria}
	\IEEEauthorblockA{\href{mailto:nils.quetschlich@tum.de}{nils.quetschlich@tum.de}\hspace{1.0cm}\href{mailto:v.koch@tum.de}{v.koch@tum.de}\hspace{1.0cm}\href{mailto:lukas.burgholzer@jku.at}{lukas.burgholzer@tum.de}\hspace{1.0cm} \href{mailto:robert.wille@tum.de}{robert.wille@tum.de}\\
	\url{https://www.cda.cit.tum.de/research/quantum}
	}

}

\maketitle
\begin{abstract}
Hundreds of satellites equipped with cameras orbit the Earth to capture images from locations for various purposes.
Since the field of view of the cameras is usually very narrow, the optics have to be adjusted and rotated between single shots of different locations. 
This is even further complicated by the fixed speed---determined by the satellite's altitude---such that the decision what locations to select for imaging becomes even more complex.
Therefore, classical algorithms for this \emph{Satellite Mission Planning Problem} (SMPP) have already been proposed decades ago.
However, corresponding classical solutions have only seen evolutionary enhancements since then.
Quantum computing and its promises, on the other hand, provide the potential for revolutionary improvement.
Therefore, in this work, we propose a hybrid classical quantum computing approach to solve the SMPP combining the advantages of quantum hardware with decades of classical optimizer development. 
Using the \emph{Variational Quantum Eigensolver}~(VQE), \emph{Quantum Approximate Optimization Algorithm}~(QAOA), and its \emph{\mbox{warm-start}} variant~(\mbox{W-QAOA}), we demonstrate the applicability of solving the SMPP for up to $21$ locations to choose from.
This \mbox{proof-of-concept}---which is available on GitHub (\url{https://github.com/cda-tum/mqt-problemsolver}) as part of the \emph{Munich Quantum Toolkit} (MQT)---showcases the potential of quantum computing in this application domain and represents a first step toward competing with classical algorithms in the future.
\end{abstract}

\section{Introduction}\label{sec:intro}

Satellites orbiting the Earth have become a critical tool for a wide range of applications, including environmental monitoring, navigation, as well as communication. 
These satellites are usually equipped with cameras to capture images of the Earth's surface.
However, this comes with a significant challenge: Satellites have limited time and resources to capture images, and decisions must be made what areas of the Earth to cover---requiring sophisticated algorithms to efficiently and effectively allocate resources and plan the routing of the satellite.

This \emph{Satellite Mission Planning Problem} (SMPP) is a computationally challenging \emph{combinatorial optimization problem} that involves scheduling satellite resources to obtain images of the Earth's surface. 
So far, the SMPP has been approached by utilizing classical algorithms---both using \emph{heuristics} (e.g., simulated annealing and tabu search~\cite{bensanna1996exact}) and exact approaches (e.g., linear integer programming~\cite{lemaitre1997daily} or \mbox{\emph{branch-and-bound}}~\cite{verfaillie1996russian}).
While these works have already been proposed decades ago, more recent papers propose extensions of the SMPP focusing on additionally objectives, e.g., considering also transmission tasks~\cite{peng2017heuristic} or multiple satellites~\cite{chong2011distributed, hao2014improved}. 
Although a wide range of classical approaches for the SMPP and related problems have been developed, their potential for further improvements is rather small---especially, because the classical computing hardware development has shifted from revolutionary enhancements decades ago toward rather small, incremental enhancements.

Quantum computing, on the contrary, has provided tremendous improvements especially in recent years---both in software and hardware. 
Meanwhile, dedicated workflows such as those proposed in~\cite{quetschlich2023mqtproblemsolver, poggel2022recommending} are available that allow solving problems that are computationally hard for classical algorithms.
Especially, hybrid classical quantum algorithms which combine classical and quantum computing have shown promising results in solving various problems, such as in finance~\cite{stamatopoulosOptionPricingUsing2020}, optimization~\cite{zoufalQuantumGenerativeAdversarial2019}, chemistry~\cite{kandalaHardwareefficientVariationalQuantum2017a}, and physics~\cite{roggeroQuantumComputingNeutrinonucleus2020}. 
Even a quantum annealing approach to solve the SMPP has already been proposed in~\cite{stollenwerk2020image}.

In this paper, we propose a hybrid classical quantum approach for the SMPP. 
Therefore, the problem is formulated as a \emph{Quadratic Unconstraint Binary Optimization} (QUBO) problem and encoded into a quantum circuit based on the \emph{Variational Quantum Eigensolver} (VQE,~\cite{peruzzoVariationalEigenvalueSolver2014b}), the \emph{Quantum Approximate Optimization Algorithm} (QAOA,~\cite{farhi2014quantum}), and its \emph{\mbox{warm-start}} variant (\mbox{W-QAOA},~\cite{Egger2021warmstartingquantum}).
The resulting circuits are then executed on a \mbox{noise-free} and a \mbox{noise-aware} quantum simulator.
Evaluations show that the proposed approach can solve the SMPP for up to $21$ possible images in reasonable time with often (\mbox{close-to}) optimal results---demonstrating the potential quantum computing offers in this application domain.
All implementations are available on GitHub (\url{https://github.com/cda-tum/mqt-problemsolver}) as part of the \emph{Munich Quantum Toolkit} (MQT).
Combined with the anticipated trend of improvements for both quantum hardware and software, this approach may become a serious competitor for the currently used classical solvers in the future.

The remainder of this work is structured as follows.
\autoref{sec:problem} reviews the SMPP in more detail and describes the hybrid classical quantum approach proposed in this work.
The applicability of the resulting solution has been confirmed by experimental evaluations whose results are summarized in \autoref{sec:eval}.
Finally, \autoref{sec:discussion} discusses the lessons learned before the paper is concluded in \autoref{sec:conclusions}.

\vspace{10cm}
\section{Satellite Mission Planning Problem: \\Hybrid Classical Quantum Approach}\label{sec:problem}
In this section, the SMPP is described and formulated based on a more comprehensive problem description proposed in~\cite{stollenwerk2020image}.
Thereafter, all the steps required to derive a quantum \mbox{computing-based} solution are described.

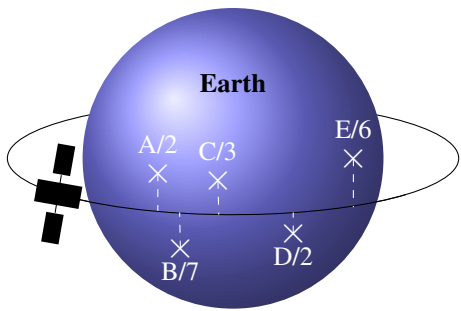
\begin{figure}[t]
\centering
\tikzset{partial ellipse/.style args =
  {#1:#2:#3}{insert path={+ (#1:#3) arc (#1:#2:#3)}}}
\begin{tikzpicture}[>=latex]
  \draw [] (3,-1.8) ellipse (3cm and 0.75 cm);

  \shade [ball color=white!50!blue] (3,-1.8) circle (2);
  \node [text=black](earth) at (3,-1.8) {};
 \node[color=black, yshift=10mm] at (earth) {\bf Earth}; 
  \draw (3,-1.8) [partial ellipse=220:320:3cm and 0.75cm];

 \node[rectangle, fill=black, minimum width=6mm, minimum height=3mm, xshift=-22mm, yshift=-3.5mm, rotate=-10] (body) at (earth.south west) {}; 
 \node[rectangle, fill=black, minimum width=1mm, minimum height=4mm, yshift=3mm, xshift=0.5mm, rotate=-10] (left) at (body.north) {}; 
 \draw[rotate=-10](left.south) -- (body.center); 
 \node[rectangle, fill=black, minimum width=1mm, minimum height=4mm, yshift=-3mm, xshift=-0.5mm, rotate=-10] (right) at (body.south) {}; 
 \draw[rotate=-10](right.north) -- (body.center);

 \node[color=white, yshift=-2mm, xshift=-10mm] (L1) at (earth) {\Large $\times$}; 
 \node[color=white, yshift=1mm] at (L1.north) { A/2}; 
 \node[color=white, yshift=-12mm, xshift=-7mm] (L2) at (earth) {\Large $\times$}; 
 \node[color=white, yshift=-6mm] at (L2.north) {B/7}; 
 \node[color=white, yshift=-3mm, xshift=-2mm] (L3) at (earth) {\Large $\times$}; 
 \node[color=white, yshift=1mm] at (L3.north) {C/3}; 
 \node[color=white, yshift=-10mm, xshift=8mm] (L4) at (earth) {\Large $\times$}; 
 \node[color=white, yshift=-6mm] at (L4.north) {D/2}; 
 \node[color=white, yshift=0mm, xshift=16mm] (L5) at (earth) {\Large $\times$}; 
 \node[color=white, yshift=1mm] at (L5.north) {E/6}; 
 
\draw[dashed, color=white] (L1.center) -- (2,-2.5);
\draw[dashed, color=white] (L2.center) -- (2.29,-2.52);
\draw[dashed, color=white] (L3.center) -- (2.81,-2.55);
\draw[dashed, color=white] (L4.center) -- (3.80,-2.52);
\draw[dashed, color=white] (L5.center) -- (4.6,-2.45);
\end{tikzpicture}
	\caption{Satellite Mission Planning Problem.}
	\label{fig:satellite_problem}
\end{figure}

\subsection{Problem Description}
The \emph{Satellite Mission Planning Problem} (SMPP) refers to the complex task of selecting a subset of feasible locations from a \mbox{pre-determined} list of locations \mbox{to-be-captured} by a satellite.
Each captured location is associated with a value. 
As a result of the narrow field of view of the satellite, its optics have to be adjusted and rotated between single shots of different locations---resulting in a \mbox{time-consuming} process referred to as \enquote{rotation time}. 
This is even further complicated by the fixed speed of the satellite (determined by the satellite's altitude) which leads to a fixed \enquote{transition time} between the \emph{capture positions} of locations (for the sake of simplification, it is assumed that each location can only be captured when the satellite is at its closest distance).

\begin{example}\label{ex:SMPP_instance}
Assume that an imaging satellite orbits the Earth at a given altitude with constant speed as depicted in \autoref{fig:satellite_problem} in black.
In this setup, five \mbox{to-be-captured} locations are given---denoted by $A$ to $E$ with their respective associated value if captured. 
The goal is to determine a feasible subset of locations that maximizes the sum of their associated values.
In this example, there are two infeasible pairs of locations ($A$/$B$ and $B$/$C$) for which the rotation time is larger than the transition time. 
Consequently, $A$ and $C$ or only $B$ can be selected in addition to $D$ and $E$, which are feasible in both cases.
Since the associated value of $B$ is higher than both $A$'s and $C$'s values combined, the optimal selection is $\{B, D, E\}$.
\end{example}

More formally, the SMPP can be described as a binary constrained optimization problem:
\begin{align}\label{formula:constrained}
&\max_{x\in\{0,1\}^n} \sum_{i=1}^{n} x_i  v_i \text{ with } 
\\
R(i, j) &\leq T(i, j) \text{ if }x_i=x_j=1 \nonumber 
\end{align}
with
\begin{itemize}
\item $x$ being the vector of binary variables $x_i$ representing whether a location $i$ has been selected for imaging at its respective capture position $s_i$ on the satellite orbit, 
\item $v_i$ being the associated value of a captured location $i$,
\item $R(i, j)$ being the required rotation time between two locations defined by the necessary satellite optics rotation time between angle $a_i$ at satellite position $s_i$ and angle $a_j$ at position $s_j$ divided by the constant optics rotation speed $v_r$: $\frac{|a_i-a_j|}{v_r}$, and
\item $T(i, j)$ being the transition time between two capture positions $s_i$ and $s_j$ divided by satellite speed $v_s$: $\frac{|s_i-s_j|}{v_s}$.
\end{itemize}

To derive a quantum computing-based solution to this problem, the general workflow proposed in~\cite{quetschlich2023mqtproblemsolver} is followed: 
First, a suitable quantum algorithm is selected based on the problem to be solved.
Subsequently, the problem must be encoded into a quantum circuit based on that selected algorithm.
This quantum circuit is then executed and the resulting histogram is decoded to extract the actual problem solution.

\subsection{Quantum Algorithm Selection}
The currently available quantum algorithms can be classified into two categories.
The first class, referred to as \emph{pure} quantum algorithms, has been proven to outperform its classical counter parts, such as Shor's~\cite{shorsalgo} for factorization, Grover's~\cite{groveralgo} for unsorted database search, as well as the \emph{Harrow-Hassadim-Lloyd} (HHL,~\cite{harrowQuantumAlgorithmLinear2009}) algorithm for solving linear systems of equations.
However, these algorithms generally assume \mbox{close-to} perfect quantum hardware which is not yet available to the necessary extent.

Hybrid classical quantum algorithms or the so-called \emph{Variational Quantum Algorithms} (VQAs) are the second class of quantum algorithms.
Prominent representatives are the \emph{Variational Quantum Eigensolver} (VQE,~\cite{peruzzoVariationalEigenvalueSolver2014b}) to determine the ground state of a given physical system and the \emph{Quantum Approximation Optimization Algorithm} (QAOA,~\cite{farhi2014quantum}) used for various optimization problems.
In contrast to pure quantum algorithms, the requirements regarding the quality of quantum hardware are less strict and, hence, these algorithms are frequently applied in the current \emph{Noisy-Intermediate-Scale-Quantum} (NISQ,~\cite{Preskill2018quantumcomputingin}) era.
The general idea behind VQAs is to encode a given problem into a parameterized quantum circuit and utilize a classical optimizer to determine the most suitable parameter values so that the resulting quantum circuit allows one to determine the solution to the given problem.

To also apply VQAs to the SMPP, the constrained binary problem must be \mbox{re-formulated} to be of the form of a \emph{Quadratic Unconstrained Binary Optimization} (QUBO) problem since that is the common input format for VQAs.
Therefore, a penalty term is introduced to substitute the condition to be satisfied in the binary constrained optimization problem formulation in \autoref{formula:constrained}.
This leads to
\begin{align}
\max_{x\in\{0,1\}^n} \sum_{i=1}^{n} x_i  v_i - p\cdot \sum_{i=1}^{n-1} \sum_{j=i}^{n }x_i x_j \cdot c(i, j)
\end{align}
with
\begin{itemize}
\item $p$ being the penalty value that enforces the constraints to be satisfied and
\item $c(i, j)$ being a \mbox{pre-calculated} Boolean variable which is $1$ if $R(i, j) > T(i, j)$ and $0$ otherwise. 
\end{itemize}

Using this QUBO formulation, VQAs can be utilized to solve the SMPP.

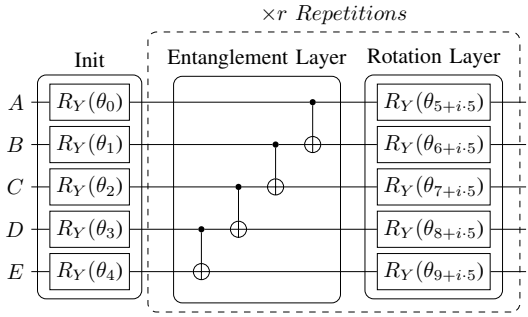
\begin{figure}[t]
\centering
   \resizebox{0.8\linewidth}{!}{
				\begin{tikzpicture}
				  \begin{yquant*}

						qubit {$A$} q;
						qubit {$B$} q[+1];
						qubit {$C$} q[+1];
						qubit {$D$} q[+1];
						qubit {$E$} q[+1];

						[this subcircuit box style={rounded corners, inner ysep=4pt,  "Init"}]
						subcircuit {
						qubit {} q[5];
				    	box {$R_Y(\theta_0$)} q[0];
				    	box {$R_Y(\theta_1$)} q[1];
				    	box {$R_Y(\theta_2$)} q[2];
				    	box {$R_Y(\theta_3$)} q[3];
				    	box {$R_Y(\theta_4$)} q[4];
						} (q[0-4]);		
					
					[this subcircuit box style={dashed, rounded corners, inner ysep=4pt,  "$\times r~Repetitions$"}]
						subcircuit {
						qubit {} q[5];
						
						[this subcircuit box style={rounded corners, inner ysep=10.5pt,  "Entanglement Layer"}]
						subcircuit {
						qubit {} q[5];
						
				    	cnot q[4] | q[3];
				    	cnot q[3] | q[2];
				    	cnot q[2] | q[1];
				    	cnot q[1] | q[0];
				    	
						} (q[0-4]);

						[this subcircuit box style={rounded corners, inner ysep=4pt,  "Rotation Layer"}]
						subcircuit {
						qubit {} q[5];
				    	box {$R_Y(\theta_{5+i\cdot5}$)} q[0];
				    	box {$R_Y(\theta_{6+i\cdot5}$)} q[1];
				    	box {$R_Y(\theta_{7+i\cdot5}$)} q[2];
				    	box {$R_Y(\theta_{8+i\cdot5}$)} q[3];
				    	box {$R_Y(\theta_{9+i\cdot5}$)} q[4];
						} (q[0-4]);		
						
						} (q[0-4]);		

				  \end{yquant*}				\end{tikzpicture}}
	\caption{SMPP encoded using VQE.}
	\label{fig:satellite_problem_encoding_vqe}
	\vspace{-4mm}
\end{figure}

\subsection{Encoding as a Quantum Circuit}
How a QUBO is translated into a quantum circuit depends on the selected quantum algorithm and the respective \emph{encoding} of the problem formulation---stating a whole research area on its own with overviews given in~\cite{dominguez2023encodingindependent, schuld2021machine, weigold2021encoding}.
For the SMPP, we used VQE, QAOA, and its \emph{\mbox{warm-start}} variant \mbox{W-QAOA} to determine a feasible subset of locations from all possible ones with each location encoded by one respective qubit.
The underlying circuit structures of those algorithms are described in the following.

The general quantum circuit structure of VQE is \mbox{problem-independent} and various ansatz functions are available. 
Here, a \emph{RealAmplitudes}\footnote{For details, see the Qiskit documentation at \url{https://qiskit.org/documentation/stubs/qiskit.circuit.library.RealAmplitudes.html}.} ansatz with reverse linear entanglement has been chosen and is depicted in \autoref{fig:satellite_problem_encoding_vqe}.
It consists of the following building blocks:
\begin{itemize}
\item Init: A layer of parameterized \mbox{single-qubit} rotations.
\item Entanglement Layer: The reverse linear entanglement ansatz is implemented based on a sequence of \emph{CNOT} gates. 
\item Rotation Layer: Every repetition ends with another layer of \mbox{single-qubit} rotations implemented using \emph{$R_Y$} gates. Each gate comes with its own rotation angle parameter.
\end{itemize}
\noindent Both the entanglement and rotation layers are repeated for $r$ repetitions while the index $i$ denotes the $i$th repetition.

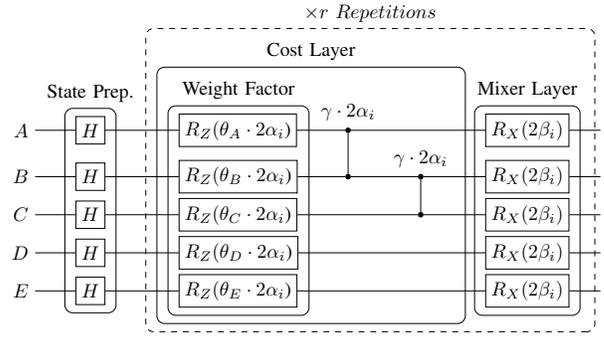
\begin{figure}[t]
\centering
   \resizebox{0.9\linewidth}{!}{
				\begin{tikzpicture}
				  \begin{yquant*}

						qubit {$A$} q;
						qubit {$B$} q[+1];
						qubit {$C$} q[+1];
						qubit {$D$} q[+1];
						qubit {$E$} q[+1];

						[this subcircuit box style={rounded corners, inner ysep=4pt,  "State Prep."}]
						subcircuit {
						qubit {} q[5];
				    	box {$H$} q[0-4];
						} (q[0-4]);		
					
					[this subcircuit box style={dashed, rounded corners, inner ysep=4pt,  "$\times r~Repetitions$"}]
						subcircuit {
						qubit {} q[5];
						
						[this subcircuit box style={rounded corners, inner ysep=4pt,  "Cost Layer"}]
						subcircuit {
						qubit {} q[5];
						
						[this subcircuit box style={rounded corners, inner ysep=4pt,  "Weight Factor"}]
						subcircuit {
						qubit {} q[5];
				    	box {$R_Z(\theta_A\cdot 2 \alpha_i$)} q[0];
				    	box {$R_Z(\theta_B\cdot 2 \alpha_i$)} q[1];
				    	box {$R_Z(\theta_C \cdot 2 \alpha_i$)} q[2];
				    	box {$R_Z(\theta_D \cdot 2 \alpha_i$)} q[3];
				    	box {$R_Z(\theta_E \cdot 2 \alpha_i$)} q[4];
						} (q[0-4]);

						[style=black]
				    	phase {$\gamma\cdot 2\alpha_i$} q[0]| q[1];
						[style=black]
				    	phase {$\gamma\cdot 2\alpha_i$} q[1]| q[2];
						} (q[0-4]);			
						
						[this subcircuit box style={rounded corners, inner ysep=4pt,  "Mixer Layer"}]
						subcircuit {
						qubit {} q[5];
				    	box {$R_X(2\beta_i)$} q[0-4];
						} (q[0-4]);		
				    	
						} (q[0-4]);

				  \end{yquant*}
				\end{tikzpicture}}
	\caption{SMPP encoded using QAOA.}
	\label{fig:satellite_problem_encoding_qaoa}
	\vspace{-4mm}
\end{figure}

The quantum circuit structure of QAOA comes with a significant difference: it is \mbox{problem-dependent} and consists of three building blocks as depicted in \autoref{fig:satellite_problem_encoding_qaoa}:
\begin{itemize}
\item State Preparation: Creates a superposition state of all qubits by applying \emph{Hadamard} gates denoted by $H$.
\item Cost Layer: Defines the SMPP instance by encoding weight factors into \emph{$R_Z$} gates and infeasible location pairs as \emph{RZZ} gates as depicted for the problem instance described in \autoref{ex:SMPP_instance}. 
Those $\theta$ and $\gamma$ factors are defined by the SMPP model describing its locations values and feasibility constraints among each other\footnote{
Note, that the $\theta$ and $\gamma$ are \mbox{pre-determined} before $\alpha_i$ and $\beta_i$ are optimized by transforming the QUBO problem into an \emph{Ising Hamiltonian}---e.g., by using respective SDKs, such as Qiskit~\cite{qiskit}.}.
\item Mixer Layer: Enables the algorithm to theoretically be able to find the optimal parameter values and, therewith, a solution to the SMPP.
\end{itemize}
\noindent Again, both the cost and the mixer layers are repeated for arbitrarily many repetitions $r$ with $i$ denoting the $i$th repetition.

The resulting quantum circuits are both now capable of determining the SMPP solution in combination with a classical optimizer which is used for tuning the parameter toward an optimal solution.
For that, their values can either be randomly initialized or, at least for QAOA, be \mbox{pre-determined} using its \emph{\mbox{warm-start}} variant (\mbox{W-QAOA},~\cite{Egger2021warmstartingquantum}).

\subsection{Execution}
While the generated quantum circuit could immediately be executed on a quantum simulator, the execution on the quantum device requires the quantum circuit to be \emph{compiled} accordingly. 
Each quantum device induces constraints on the to-be-executed quantum circuit such as on the supported elementary quantum gates and, often, additionally on the connectivity of qubits.
Various software tools have been proposed to automate the compilation in, e.g.,~\cite{quetschlich2023prediction, quetschlich2023compileroptimization}.
Thereafter, the quantum circuit is ready to be executed---resulting in a distribution of outcomes, also called a histogram, which comprises the determined solution.
To extract it, the histogram must be decoded so that the actual solution is extracted.
Although this step marks the end of the workflow for pure quantum algorithms, this differs for VQAs.

\begin{figure}[h!]
\centering
   \resizebox{0.48\linewidth}{!}{
				\begin{tikzpicture}
				  \begin{yquant*}

						qubit {$A$} q;
						qubit {$B$} q[+1];
						qubit {$C$} q[+1];
						qubit {$D$} q[+1];
						qubit {$E$} q[+1];
				    	box {Parameterized Quantum Circuit \\with Parameters $\omega$} (-);	
				  \end{yquant*}
				\end{tikzpicture}}
				   \resizebox{0.48\linewidth}{!}{
				\begin{tikzpicture}
				  \draw (1,1) node[text width=2cm, draw=black, align=center] (opt)   {\small Classical Optimizer};
\draw[bend left,<-, yshift=2mm]  (-3,1) to node [auto] {\small $\omega'$} ([xshift=-2mm, yshift=2mm]opt.west);
\draw[bend right,->, yshift=-2mm]  (-3,1) to node [auto, below] {\small Cost($\omega$)} ([xshift=-2mm, yshift=-2mm]opt.west);
				\end{tikzpicture}}
	\caption{VQA execution scheme.}
	\label{fig:vqa_scheme}
\vspace{-2mm}
\end{figure}
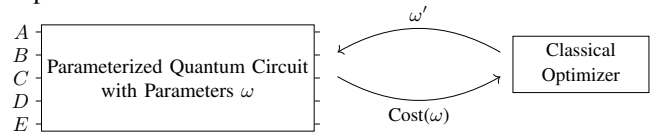

Here, as sketched in \autoref{fig:vqa_scheme}, the decoded solution is used to determine a \emph{Cost} value which depends on the chosen parameter values of the parameterized quantum circuit and the problem itself.
\vspace{-5mm}

\newcommand{\picwidth}{0.48}
\begin{figure*}
     \centering
     \begin{subfigure}[b]{\picwidth\linewidth}
         \centering
         \includegraphics[width=\textwidth]{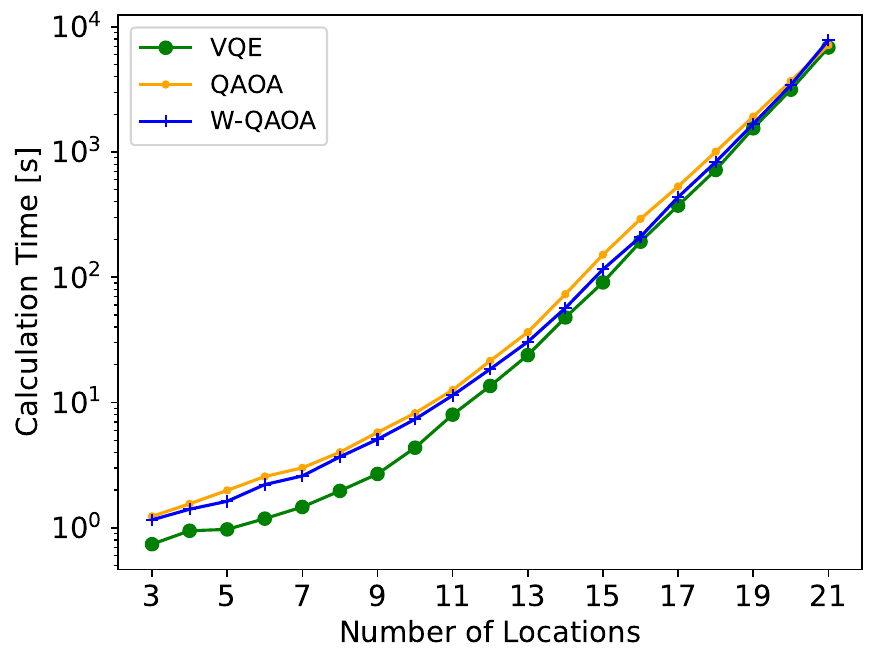}
         \caption{Calculation time for SMPP instances.}
         \label{fig:res_time}
     \end{subfigure}
     \hfill
     \begin{subfigure}[b]{\picwidth\linewidth}
         \centering
         \includegraphics[width=\textwidth]{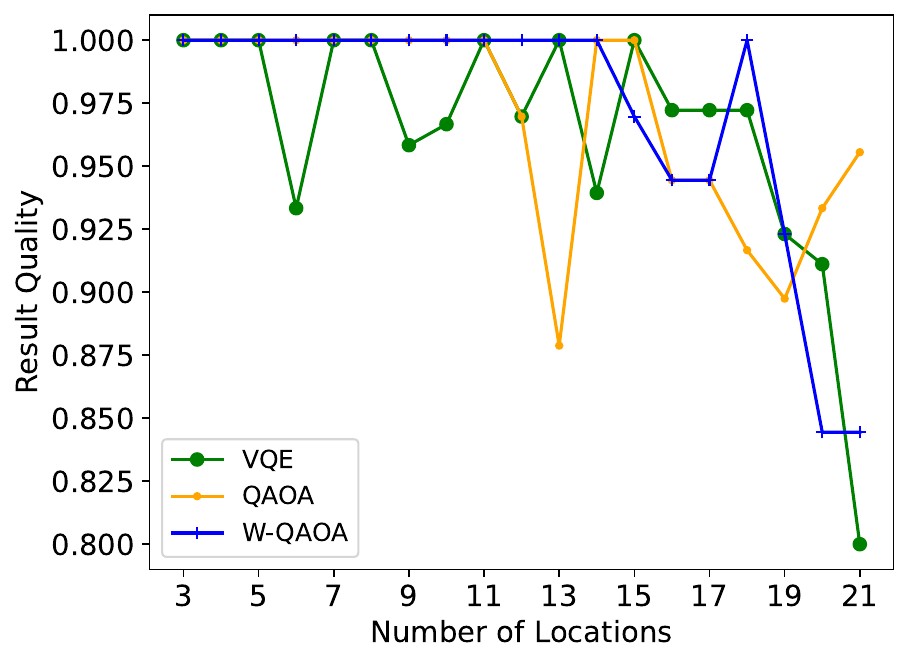}
         \caption{Result quality for SMPP instances.}
         \label{fig:res_quality}
     \end{subfigure}
\caption{Experimental evaluation for the satellite use-case using a noise-free simulation.}
\label{fig:results}
\end{figure*}

This is also what enables VQE to determine solutions while the problem is not represented in the quantum circuit itself.
The classical optimizer then adjusts the parameter values to evaluate their influence in the next iteration. 
With many of such iterations, the parameter values ideally converge to their optimal value.
When this is the case, the algorithm terminates and the decoded solution of the quantum circuit executed with the optimal parameters is returned.

\section{Experimental Evaluation}\label{sec:eval}
In the following, the applicability of quantum computing to the SMPP is demonstrated.
To this end, we first describe the evaluation setup and the problem instances considered.
Subsequently, the results based on \mbox{noise-free} and \mbox{noise-aware} simulations are presented and discussed.

\subsection{Setup}
All implementations have been carried out using Qiskit~\cite{qiskit} (v$0.41.1$) in \emph{Python} and are available on GitHub~(\url{https://github.com/cda-tum/mqt-problemsolver}) as part of the \emph{Munich Quantum Toolkit}~(MQT).
To evaluate the proposed approach, SMPP instances from $3$ up to $21$ \mbox{to-be-captured} locations have been solved using VQE, QAOA, and W-QAOA in a \mbox{noise-free} simulation and up to $14$ locatins in a \mbox{noise-aware} one.
The problem instances themselves are generated as follows: Each image location is randomly placed with an arbitrary longitude while the latitude is restricted to be between $-15$ and $15$ degrees (since the satellite is assumed to orbit around the equator and this condition ensures that each location is visible to the satellite) and is assigned a value of either $1$ or $2$.

For both, the \mbox{noise-free} and the \mbox{noise-aware} simulation, \emph{Cobyla} is used as the classical optimizer.
The maximal iterations are set to $100$ for both optimizers in all evaluations.
All VQAs are evaluated using $3$ repetitions each.
For VQE, a \emph{RealAmplitudes} ansatz with reverse linear entanglement has been chosen.

To evaluate both the \mbox{noise-free} and the \mbox{noise-aware} simulations, the calculation time and result quality are evaluated.

\begin{figure*}
     \centering
     \begin{subfigure}[b]{\picwidth\linewidth}
         \centering
         \includegraphics[width=\textwidth]{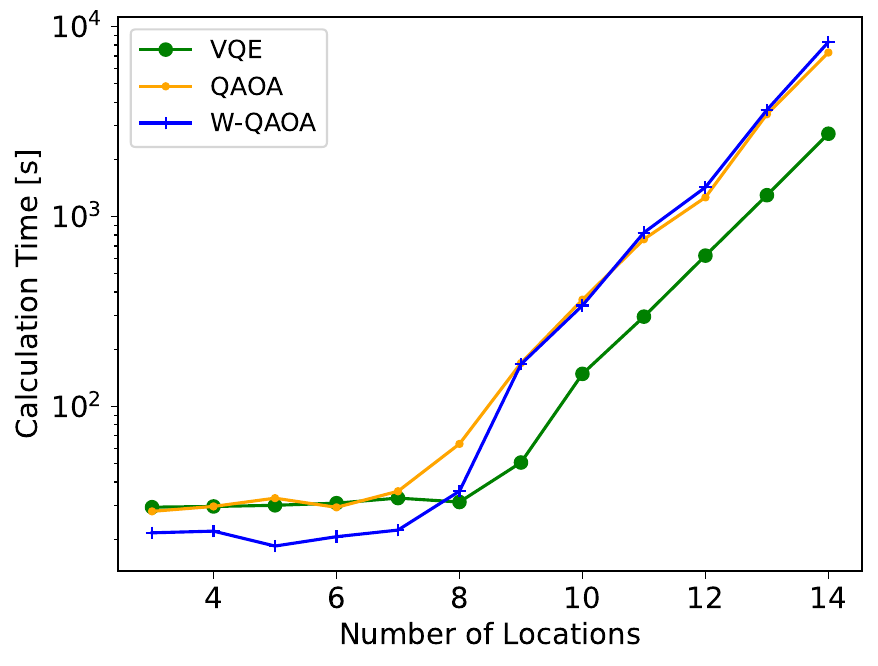}
         \caption{Calculation time for SMPP instances.}
         \label{fig:res_time_noisy}
     \end{subfigure}
     \hfill
     \begin{subfigure}[b]{\picwidth\linewidth}
         \centering
         \includegraphics[width=\textwidth]{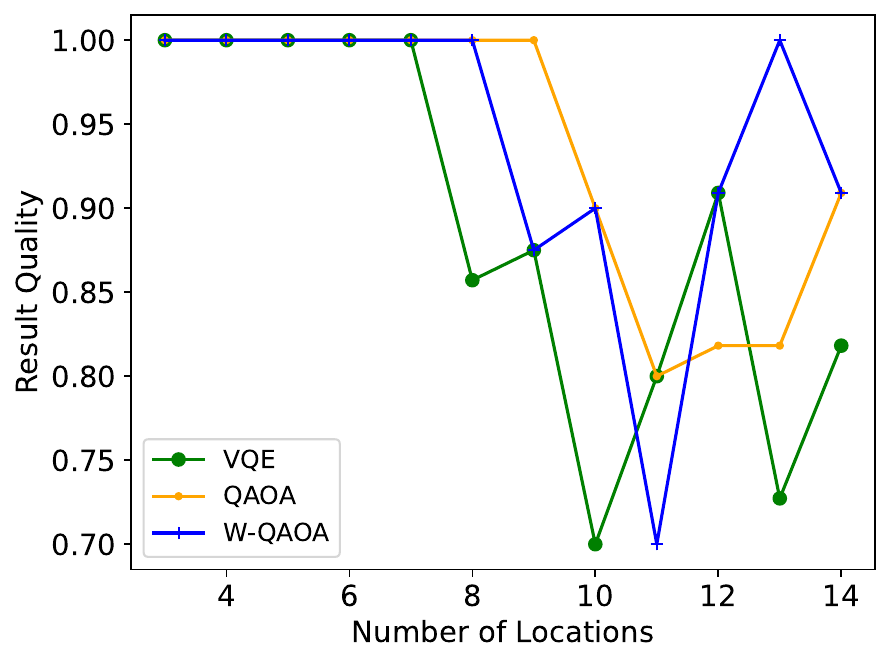}
         \caption{Result quality for SMPP instances.}
         \label{fig:res_quality_noisy}
     \end{subfigure}
\caption{Experimental evaluation for the satellite use-case using a \mbox{noise-aware} simulation.}
\label{fig:noisy_res}
\vspace{-6mm}
\end{figure*}

\subsection{Noise-free Simulation}
Each problem instance is executed three times per algorithm and both the calculation time and the result quality are averaged to be more representative.
All simulations are conducted using Qiskit's \mbox{matrix-vector-based} \emph{Sampler} simulator.
The resulting computation times for all three algorithms are provided in \autoref{fig:res_time}.
As expected, the computation time quickly increases with the number of locations and shows an exponential trend (represented by the linear slope on a logarithmic scale) while VQE was significantly faster for smaller problem instances.

The respective result quality is shown in \autoref{fig:res_quality}.
To this end, a classical computation of the optimal solution and the corresponding value is performed, which is then used as a baseline to measure the relative deviation of the VQAs---e.g., a result quality of $0.95$ corresponds to a location selection which leads to an assigned accumulated value of $95\%$ of the optimal one.

In general, there is a clear trend of degrading result quality with an increasing number of locations.
Until around $12$ locations, all algorithms have been able to determine the (\mbox{close-to}) optimal results with a result quality of $>90\%$.
From there on, a notable decrease in result quality is occurring with VQE degrading the most---down to a result quality of $0.80$ for the largest problem instance size of $21$ locations.

\subsection{Noise-aware Simulation}
The \mbox{noise-aware} simulation is based on Qiskit's \mbox{\emph{BackendSampler}} simulator using the \emph{FakeBackend} for \emph{ibmq\_montreal} with $27$ qubits.  
The respective computation times for all algorithms are provided in \autoref{fig:res_time_noisy} and follow a similar exponential trend for an increasing number of locations with \mbox{W-QAOA} being the fastest algorithm for smaller problem instances and VQE for larger ones.
Compared to the \mbox{noise-free} simulation, the computation times are significantly higher for similar problem sizes.
This is caused by a combination of three factors: the classical optimizer requires more iterations to converge at each execution, the additional compilation time to compile the underlying quantum circuit such that it is executable on the selected (but simulated) quantum device, and the increased simulation time for a \mbox{noise-aware} simulation.
Due to this increase in computation time, the maximal number of locations evaluated is smaller compared to the \mbox{noise-free} simulation.

\vspace{10cm}

The result quality is shown in \autoref{fig:res_quality_noisy} with a qualitatively similar trend of decreasing result quality for an increasing number of locations.
However, the quality drop is greater with the lowest quality around $70\%$---already for a number of locations where the result quality was \mbox{close-to} optimal in the \mbox{noise-free} simulation.

Although the literature suggests that one uses the \emph{SPSA} optimizer for \mbox{noise-aware} simulations (e.g., in~\cite{cumming2022using}), that optimizer led to worse results while taking more computation time.
The reason could be that the \emph{loss landscapes} and the respective solutions of the considered SMPPs have been sufficiently distinct for the faster \emph{Cobyla} optimizer not to degrade too much due to the induced noise.

Through the studies summarized above, the potential for solving the SMPP using VQAs has been evaluated for the first time.
Based on that, the \enquote{lessons learned} are discussed in the following.

\section{Discussion}\label{sec:discussion}
VQAs have demonstrated promising results in solving combinatorial optimization problems, including those in the satellite domain. 
However, they face several challenges that need to be addressed. 
One significant limitation is the requirement for a large number of qubits to solve larger problems (since each \mbox{to-be-captured} location is encoded as one qubit)---easily outgrowing what current simulators can reasonably simulate and the capabilities of available quantum devices.
Furthermore, VQAs are known to suffer from the issue of barren plateaus, which can make finding the optimal solution difficult, particularly in complex optimization problems. 
Determining the optimal set of parameters for the classical optimizer is another challenge, particularly for large problems, and can increase the computational cost and time required for solving optimization problems.

Although classical solvers currently outperform VQAs, these hybrid classical quantum algorithms, in general, still present a promising approach since the ongoing development of quantum hardware and software is expected to improve the performance of VQAs---making it an increasingly viable option for solving optimization problems in the future.

\section{Conclusions}\label{sec:conclusions}
Due to the increasing number of satellites and their limited resources, the \emph{Satellite Mission Planning Problem}~(SMPP) becomes more relevant.
Classical solutions have already been proposed decades ago but have only seen evolutionary enhancements since then.
Quantum computing, on the other hand, has the potential for a revolutionary improvement.
To this end, a hybrid classical quantum computing approach to the SMPP is proposed in this work using three different \emph{Variational Quantum Algorithms}: VQE, QAOA, and \mbox{W-QAOA}.
To demonstrate their potential, SMPP instances have been experimentally evaluated in both a \mbox{noise-free} with up to $21$ \mbox{to-be-captured} locations and a {noise-aware} simulation with up to $14$ \mbox{to-be-captured} locations---often leading to (\mbox{close-to}) optimal results.
This \mbox{proof-of-concept} showcases the general applicability of quantum computing to solve the SMPP and is available on GitHub (\url{https://github.com/cda-tum/mqt-problemsolver}) as part of the \emph{Munich Quantum Toolkit} (MQT).
Although the proposed quantum approach cannot compete with the classical solutions yet, promising developments in both quantum hardware and software---especially in recent years---might change that in the future.

\section*{Acknowledgments}
This work received funding from the European Research Council (ERC) under the European Union’s Horizon 2020 research and innovation program (grant agreement No. 101001318), was part of the Munich Quantum Valley, which is supported by the Bavarian state government with funds from the Hightech Agenda Bayern Plus, and has been supported by the BMWK on the basis of a decision by the German Bundestag through project QuaST, as well as by the BMK, BMDW, and the State of Upper Austria in the frame of the COMET program (managed by the FFG).
Finally, we thank the organizers of the Quantum Entrepreneurship Lab which sparked the idea for this work.

\vspace{10cm}
\clearpage
\printbibliography

\end{document}